\documentstyle[12pt,twoside,fleqn,espcrc1]{article}

\def\tit{Department of Physics,
        Tokyo Institute of Technology, 
        Tokyo 152-8551, Japan}

\newcommand{\AmS}{{\protect\the\textfont2
  A\kern-.1667em\lower.5ex\hbox{M}\kern-.125emS}}

\title{ QCD sum rules with two-point correlation function
}

\author{
Hungchong Kim\address{\tit} \thanks{JSPS Fellow.},
Su Houng Lee\address{GSI, Planckstr. 1, D-64291 Darmstadt,
Germany}\address{Department of Physics, Yonsei University,
Seoul 120-749, Korea}\thanks{AvH Fellow.}
and Makoto Oka\addtocounter{address}{-3}\addressmark
}

\begin{document}
\maketitle

\begin{abstract}
We construct three different sum rules from the
two-point correlation function with pion,
$i\int d^4x e^{iq\cdot x} \langle 0| T J_N(x) {\bar J}_N(0)|\pi(p)\rangle$,
beyond the soft-pion limit.  The PS and PV coupling schemes in the
construction of the phenomenological side are carefully considered in 
each sum rule. 
We discuss the dependence of the result on the specific Dirac 
structure and identify the source of the dependence
by making specific models for higher resonances. 
\end{abstract}

\vspace{15pt}


Within QCD sum rules, the $\pi NN$ coupling constant, $g_{\pi N}$, is 
often calculated~\cite{hat,krippa}, for example, from the correlation function,
\begin{eqnarray}
i\int d^4x e^{iq\cdot x} \langle 0| T J_p(x) {\bar J}_n(0)|\pi^+ (p)\rangle\ ,
\label{corr}
\end{eqnarray}
where $J_p$ is the proton interpolating field~\cite{ioffe1} 
and $J_n$ is the neutron interpolating field.
Shiomi and Hatsuda~\cite{hat} considered $i\gamma_5$ Dirac structure 
from this correlation function in the soft-pion limit 
($p_\mu \rightarrow 0$). Later, Birse and Krippa~\cite{krippa}
pointed out that the use of soft-pion limit does not constitute an
independent sum rule from the nucleon chiral-odd sum rule, and proposed
to look at the Dirac structure, $i\gamma_5 \not\!p$, beyond the
soft-pion limit.  

Recently~\cite{hung}, we have pointed out that
the previous calculations of this sort have dependence on how 
one models the phenomenological side; either using the 
pseudoscalar (PS) or the 
pseudovector (PV) coupling scheme. Beyond the
soft-pion limit,  we presented a new sum rule 
for the 
$\gamma_5 \sigma_{\mu \nu} {q^\mu p^\nu}$ structure. 
This sum rule  is independent of the coupling schemes and
provides $g_{\pi N}$ relatively close to its empirical value.

Then we ask, can we get similar results
from the other Dirac structures, $i\gamma_5$ and $i\gamma_5 \not\!p$,
constructed beyond the soft-pion limit ?  
If not, what are the
reasons for the differences ?
In this work, we will try to answer these questions
by studying all three sum rules  and
investigating the reliability of each sum rule.

In calculating the OPE of Eq.~(\ref{corr}),  we only keep the 
quark-antiquark component of the pion wave function,
\begin{eqnarray}
D^{\alpha\beta}_{a a'} \equiv 
\langle 0 | u^\alpha_a (x) {\bar d}^\beta_{a'} (0) | \pi^+ (p) \rangle\ ,
\label{qantiq}
\end{eqnarray}
and use the vacuum saturation hypothesis to factor out 
higher dimensional operators in terms of the pion wave function and the 
vacuum expectation value. This quark-antiquark component can be
written in terms of the following three matrix elements,
\begin{eqnarray}
\langle 0 |  
{\bar d} (0) \gamma_\mu \gamma_5  u (x) | \pi^+ (p) \rangle\;, \quad \ 
\langle 0 |  
{\bar d}(0) \gamma_5 \sigma_{\mu\nu}  u (x) | \pi^+ (p) \rangle\;, \quad \
\langle 0 |  
{\bar d}(0) i \gamma_5  u (x) | \pi^+ (p) \rangle\ ,  
\label{dd2}
\end{eqnarray}
whose few moments are relatively well-known~\cite{bely}.

The first two matrix elements participate in 
the sum rules with the Dirac structures,
$i\gamma_5\not\!p$ and $\gamma_5 \sigma_{\mu\nu} q^\mu p^\nu$,
while the last matrix element participates only in 
the $i \gamma_5$ sum rule. Following the standard prescription of QCD 
sum rule, we obtain for the $i\gamma_5\not\!p$ structure,
\begin{eqnarray}
&&g_{\pi N} \lambda^2_N (1 + A M^2) \nonumber \\ 
&&= {f_\pi \over m} M^2 e^{m^2/M^2} \left [
{E_1 (x_\pi) \over 2 \pi^2} M^4 + 
{E_0 (x_\pi) \over 2 \pi^2} M^2 \delta^2
+ {1 \over 12} 
\left \langle {\alpha_s \over \pi} {\cal G}^2 
\right \rangle  +
{2 \langle {\bar q} q \rangle^2 \over 9 f_\pi^2 } \right ]\ .
\label{bksum}
\end{eqnarray}
Here $x_\pi = S_\pi/M^2$ with $S_\pi$ being the continuum threshold and 
$E_n (x) = 1 -(1+x+ \cdot \cdot \cdot + x^n/n!)~ e^{-x}$ .
$\delta^2$ is the twist-4 contribution to the first matrix element in 
Eq.~(\ref{dd2}).
The unknown single pole, $A$, contains the contribution of 
$N \rightarrow N^*$~\cite{ioffe2} as well as the  
PS-PV scheme dependent $N \rightarrow N$~\cite{hung}.  Thus, physical
content of $A$ is coupling-scheme dependent.  
In obtaining Eq.~(\ref{bksum}), we have taken out one power of the pion
momentum and took the limit
$p_\mu \rightarrow 0$ in the rest of the correlator.
The corresponding expression in Ref.~\cite{krippa}, not only missing the last
term in Eq.~(\ref{bksum}),   
contains
different continuum factors, $E_i(x_\pi)$.  The sum rule result
strongly depends on these~\cite{hung1}.

The sum rule for $\gamma_5 \sigma_{\mu\nu} q^\mu p^\nu$ can be 
constructed similarly~\cite{hung},
\begin{eqnarray}
g_{\pi N} \lambda_N^2  ( 1+ B M^2  )= 
- {\langle {\bar q} q \rangle \over 
f_\pi} e^{m^2/M^2} \left [ {M^4 E_0 (x_\pi) \over 12 \pi^2 }
+{4 \over 3 } f^2_\pi M^2  + 
\left \langle {\alpha_s \over \pi} {\cal G}^2
\right \rangle 
{1 \over 216 } 
-{m_0^2 f^2_\pi \over 6 }
\right ]\ .
\label{hungsum}
\end{eqnarray}
The unknown single pole term is represented by $B$ whose physical content is
independent of the coupling schemes.  This can be also checked explicitly by 
constructing $B$ using effective models for higher resonances. 
$m_0^2$ is the parameter associated with the dim-5 quark-gluon
mixed condensate.
We emphasize that this sum rule is independent of the PS and PV coupling 
scheme employed in the phenomenological side.   When this sum rule is 
combined with the
nucleon chiral odd sum rule, it yields $g_{\pi N} \sim 10$~\cite{hung}.

The sum rule for the $i \gamma_5$ structure beyond the soft-pion
limit is complicated due to the PS or PV coupling scheme dependence.
By expanding the correlator in terms of the pion momentum, one can
construct various sum rules for this structure at each order 
of the pion momentum. But none of them
is independent of the coupling schemes.   
To achieve the coupling scheme independence, we  
introduce the kinematical condition,
\begin{equation}
p^2 = 2 p \cdot q\label{kine} ,
\end{equation}
which places each nucleon on its mass-shell. 
In this sum rule, only the last term in Eq.~(\ref{dd2}) contributes to 
the OPE side.
By keeping up to the order $p^2$ in the expansion of
the correlator in terms of the pion momentum,
we obtain,
\begin{eqnarray}
g_{\pi N} \lambda^2_N (1 + C M^2) = 
{\langle {\bar q} q \rangle 
\over f_\pi} e^{m^2/M^2} \left [
{0.0785 E_0 (x_\pi) \over \pi^2} M^4
- 0.314\times {1 \over 24} 
\left \langle {\alpha_s \over \pi} {\cal G}^2 
\right \rangle  \right ]\ .
\label{shsum}
\end{eqnarray}
Again $C$ denotes the unknown single pole term.
Note that this sum rule contains very small
numerical factors in the RHS.  This is because two independent 
correlators cancel each other when they are combined via Eq.~(\ref{kine}).

The LHSs of the three sum rules, Eqs.~(\ref{bksum}), (\ref{hungsum}) and
(\ref{shsum}), can be written as $c + b M^2$ where $c$ in all sum rules
represents $g_{\pi N} \lambda_N^2$ and $b$ denotes the unknown single 
pole terms whose physical content can be different in each sum rule.  
We determine $b$ and $c$ by linearly fitting the
RHSs within the relevant Borel window and list them in Table 1.

\begin{table}[hbt]
\newlength{\digitwidth} \settowidth{\digitwidth}{\rm 0}
\catcode`?=\active \def?{\kern\digitwidth}
\caption{
The best-fitted values for the parameters $c$ and $b$ obtained within the
Borel window $0.8 \le M^2 \le 1.2$ GeV$^2$. The continuum threshold
$S_{\pi} = 2.07$ GeV$^2$ is used.  To see the sensitivity to
the continuum, the results with $S_{\pi}=2.57 $ GeV$^2$ are
also presented in parenthesis.
The third column
is the difference of the two numbers in the second column.}
\begin{tabular*}{\textwidth}{@{}l@{\extracolsep{\fill}}rrrr}
\hline
               & {$c$ (GeV$^6$)} & {$|\Delta c|$ (GeV$^6$)} &{$b$ (GeV$^4$)} \\ 
\hline
{$i\gamma_5\not\!p$} & {-0.00022 (-0.0023)} & {0.0021}& {0.011 (0.0145)} \\
{$i\gamma_5$} & {-0.00033 (-0.00016)}& {0.00017} & {-0.00183 (-0.0021)}  \\
{$\gamma_5 \sigma_{\mu\nu} q^\mu p^\nu$} & {0.00308 (0.002906)}&{0.00017} & 
{0.00257 (0.0029)} \\ 
\hline 

\end{tabular*}
\end{table}

As shown, the extracted value of $c = g_{\pi N} \lambda_N^2$ is 
quite different depending
on Dirac structures.  The $i\gamma_5 \not\!p$ sum rule not only
contains the large single pole term (i.e. large $b$) but is also
quite sensitive to the continuum threshold (i.e. large $|\Delta c|$).
Therefore, its prediction for $c$ contains large uncertainty due
to these effective parameters. 
The other two sum rules, even though they yield  quite different value for $c$,
 contain relatively small contribution from
the unknown single pole and is less sensitive to the continuum.

The difference in the sensitivity to the continuum threshold as well 
as in the
magnitude of $b$ can be understood by making effective models
for the continuum and the unknown single pole~\cite{hung1}. 
Specifically, using the effective Largrangians for $N \rightarrow N^*$ and
$N^* \rightarrow N^*$, we identify the terms corresponding to the
unknown single pole and the step-like continuum.  There are two ways
to construct the Lagrangians, nonderivative coupling scheme and derivative
coupling scheme.
It turns out that the $i\gamma_5$ structure within the kinematical condition
of Eq.~(\ref{kine}) takes the same form as the 
$\gamma_5 \sigma_{\mu\nu} q^\mu p^\nu$ structure.  This explains the
similarities between these sum rules.  
The smallness of $b$ and $|\Delta c|$ 
can be understood by the cancellation between the positive- and negative-
parity higher resonances. This explanation is independent of the
coupling schemes.

The large $b$ and $|\Delta c|$ for the $i\gamma_5 \not\!p$ sum rule
within the nonderivative coupling scheme can be understood by adding up
contributions from the different parity resonances.   
On the other hand, in the case of the derivative coupling scheme, 
the additional single pole
of $N \rightarrow N$~\cite{hung} can explain the large $b$.
In this case, explanation for the strong sensitivity to the continuum   
threshold is not unique.  

Then, why do the $i\gamma_5$ and $\gamma_5 \sigma_{\mu\nu} q^\mu p^\nu$ 
sum rules lead to different values for  $c$ even though
they share similar features for the continuum and
the unknown single pole ? This is an interesting question to pursue in 
future.  
At this stage, it is not clear if the difference
between the two sum rules  is due to the lack of convergence
in the OPE or due to the limitations in the sum rule method itself.
In future, it will be interesting to study $i\gamma_5$ sum rule
in more detail
without imposing Eq.~(\ref{kine}). Then the sum rule results clearly
depend on the choice of the PS and PV coupling schemes, which
can provide further insights into the pion-nucleon coupling.

Nevertheless, our study in this work, though it is specific to the
two-point correlation function with pion, raises important issues in applying
QCD sum rules in calculating various physical quantities.  
Sum rules results could have strong dependence on the specific Dirac
structure one considers.  Similar issue has been raised in Ref.~\cite{jin}
for the case of the nucleon sum rule.  Anyway,  according to our study,
the $i\gamma_5 \not\!p$ structure does not constitute a reliable sum rule
as its results are contaminated by the two phenomenological inputs, the
unknown single pole and the continuum threshold.

\section*{Acknowledgments}

This work is supported in part by the
Grant-in-Aid for JSPS fellow, and
the Grant-in-Aid for scientific
research (C) (2) 08640356
of  the Ministry of Education, Science, Sports and Culture of Japan.
The work of  H. Kim is also supported by Research Fellowships of
the Japan Society for the Promotion of Science.
The work of S. H. Lee is supported by  KOSEF through grant no. 971-0204-017-2
and 976-0200-002-2 and  by the
Korean Ministry of Education through grant no. 98-015-D00061.

\end{document}